\documentstyle[11pt]{article}
\begin{document}
\bibliographystyle{unsrt}
\newpage
\thispagestyle{empty}
{\mbox{ }}\\
{\vspace{-4 cm}}{\mbox{ }}\\
{\mbox{ }}{\hspace{10 cm}}UPR-0540-T\\
{\mbox{ }}\\
{\mbox{ }}\\
{\mbox{ }}\\
{\mbox{ }}\\
{\mbox{ }}\\
\begin{center}
{\LARGE{\bf Inverse square law of gravitation in
(2+1)-dimensional
space-time
as a consequence
 of Casimir energy}}\\
{\mbox{ }}\\
{\mbox{ }}\\
\renewcommand{\thefootnote}{\fnsymbol{footnote}}
Harald H.\ Soleng\footnote{On leave from
the University of Oslo, Norway}\footnote{Email address:
HARALD@nordita.dk
}\footnote{Present address:
NORDITA, Blegdamsvej 17, DK-2100 Copenhagen {\O}, Denmark}\\
{\em Department of Physics\\
University of Pennsylvania\\
209 South 33rd Street\\
Philadelphia, Pennsylvania 19104-6396\\
 U.\ S.\ A.}

\end{center}
{\mbox{ }}\\

\begin{quote}
{\bf Abstract:}
The gravitational  effect of vacuum polarization
in space exterior to
a particle in (2+1)-dimensional
Einstein theory is investigated.
In the weak field limit this gravitational
field corresponds to an inverse
square law of gravitational attraction, even though
the gravitational mass of the quantum vacuum is negative.
The paradox is resolved by considering a particle of
finite extension and taking into account
the vacuum polarization in
its interior.
\end{quote}

\begin{center}
{\mbox{ }}\\
(April 1993)\\
{\mbox{ }}\\
{\em To appear in Physica Scripta}
\end{center}

\newpage
\addtocounter{footnote}{-\value{footnote}}
\section{Introduction}
Even though space-time is the arena of all physics, space-time
by itself
is not
fully understood. The main remaining theoretical
problem is the unification of General Relativity with quantum theory.
A possible way to reach a deeper understanding of the problem and maybe
find the key to quantum gravity is to study simpler theories that
have some of the characteristics of General Relativity. The present study
is meant to resolve a paradox encountered in semi-classical (2+1)-dimensional
Einstein theory, namely the problem that the Casimir energy outside
a point particle gives rise to an
attractive gravitational field even though the
gravitational mass of this field is negative.
For this purpose a general expression for the (2+1)-dimensional Tolman mass
is derived.

The  metric is algebraically determined by the energy-momentum tensor in
(2+1)-dimensional Einstein theory [1-5].
On account of this, space-time is flat
in the empty classical vacuum,  and there is no
Newtonian limit to Einstein's theory in `flatland'
\cite{Giddinsetal}.

Yet gravitation does exist also in empty
two dimensional space, not in terms of local curvature
but in terms of
global topological effects.
Thus, space outside a static point particle is characterized
by an   angle deficit of an otherwise Minkowskian geometry
[1-5]. A static dust planet
has the same exterior geometry \cite{GottAlpert}.

This nontrivial geometry induces  non-vanishing
vacuum expectation values of the energy-momentum tensor \cite{Dowker}
of quantum fields.
In a
semi-classical approach one should take this energy density into account
when
calculating the gravitational field.
This problem has recently been addressed by Souradeep and Sahni \cite{SS}
who computed the vacuum expectation value of the energy-momentum tensor
of massless fields exterior to a point mass and derived the gravitational
backreaction induced by this energy density.

Here we find an exact solution to Einstein's field
equations for an energy-momentum tensor of the
same algebraic form as the one induced by
massless conformally coupled quantum fields.
It is pointed out that the backreaction calculation for
a point mass gives
a paradoxical result; there is a negative gravitational mass
outside the conical singularity but still the gravitational field
corresponds to gravitational attraction.
By using Tolman's \cite{Tolman}
method of effective gravitational mass,
and deriving a formula for gravitational mass in (2+1)-dimensional
gravity,
the paradox is resolved by assuming an extended particle and by
taking into account vacuum polarization also in the interior of the
particle.

\section{The gravitational field of the quantum vacuum}

Let a static rotational symmetric space-time be described by
the space-time metric
\begin{equation}
ds^2=-A^2 dt^2+B^2 dr^2 + C^2 r^2 d\theta^2 \label{metrikk}
\end{equation}
where
$A$, $B$ and $C$ are functions of the radial coordinate $r$ only.
The solution for   classical vacuum exterior to a particle
is $A=B=1$ and $C=C_{0}\equiv 1-G\mu$, where $\mu$ is the
mass of the particle and where $G$ is the (2+1)-dimensional
Newton's constant.\footnote{For later convenience the (2+1)-dimensional
Einstein constant is chosen to be $2\pi G$,
where $G$ is a coupling constant of dimension $1/$mass.}
We shall employ units of measurements so that the speed of light is
unity.
The deviation of $C$ from unity
describes the angle deficit of the cone geometry.
By a suitable redefinition of the radial coordinate we may
 eliminate one of the metric coefficients $B$ or $C$.
 Here we choose to fix $C$ to its classical value, $C_{0}$.

The conical  topology induces a nonvanishing
vacuum expectation value of the
energy momentum tensor.
Its form may be found
by symmetry arguments. In odd-dimensional space-times there is no
conformal anomaly \cite{BD}. Besides, the background geometry is flat.
Hence, the trace of the energy-momentum tensor
of a massless conformally coupled scalar vanishes.
Then, conformal invariance and conservation of the energy-momentum tensor in
the {\em classical background\/} metric ($A=B=1$) gives
an energy-momentum tensor of the form
\begin{equation}
\langle T^{\mu}_{\;\;\nu}\rangle = \frac{a}{r^3}{\mbox{diag}}[1,1,-2]
\label{vacuum}
\end{equation}
where the coordinates are $x^{\mu}\in\{ t,$ $r,$ $\theta \}$ and
$a$ is a constant.

By using this energy-momentum tensor as a source in
Einstein's field equations
we find the quantum corrections to the classical cone geometry.
With the metric on the form (\ref{metrikk})
and $C$ constant,
the Einstein tensor has the following non-vanishing components
\begin{eqnarray}
G^{t}_{\;\;t}&=&-\frac{B'}{Br} \frac{1}{B^2} ,\label{ett}\\
G^{r}_{\;\;r}&=&\frac{A'}{Ar}\frac{1}{ B^2} ,\label{err}\\
G^{\theta}_{\;\;\theta}&=&\left[\frac{A''}{A}-\frac{A'B'}{AB}\right]
\frac{1}{B^2} \label{ethth} .
\end{eqnarray}

 With an energy-momentum tensor of the algebraic form of (\ref{vacuum}),
the semi-classical field equations
\begin{equation}
G^{\mu}_{\;\;\nu}=2\pi G\langle
T^{\mu}_{\;\;\nu}\rangle
\end{equation}
are reduced to
the
following two field equations
\begin{equation}
G^{t}_{\;\;t}=G^{r}_{\;\;r}\;\;\;{\mbox{ and }}\;\;\;
2G^{t}_{\;\;t}=-G^{\theta}_{\;\;\theta} .
\end{equation}
By introducing the new variable $u$ defined by
$ u \equiv rA'/A $
one obtains the following set of field equations
\begin{eqnarray}
{\mbox{d}}u&=&-u(2u+1)\,{\mbox{d}}r/r ,\\
{\mbox{d}}A&=&Au\, {\mbox{d}}r/r ,\\
{\mbox{d}}B&=&-Bu\,{\mbox{d}}r/r .
\end{eqnarray}
There are two solutions with constant $u$.
The first one is the Minkowski solution $u=0$. The second
solution corresponding to $u=-1/2$,
 is a power law solution for $A$ and $B$ as functions of $r$.
This is a solution with a positive energy density,
which does not smoothly reduce to the classical solution
in the limit $T^{\mu}_{\;\;\nu}\rightarrow 0$,
and hence it is not the
solution sought  here. Having thus
eliminated the solutions with constant $u$, we
may use $u$ as a new radial variable instead of $r$. Hence, integrating
$A$ and $B$ as functions of $u$ we get
\begin{equation}
A= (1+2u)^{-1/2}\;\;\; {\mbox{ and }}\;\;\;
B=(1+2u)^{1/2} .
\end{equation}
Two integration constants have been determined
 by demanding that
the metric asymptotically approaches the classical one.
The solution for $r$ is
\begin{equation}
r=\frac{1+2u}{u}\ell ,  \label{req}
\end{equation}
where $\ell$ is an aribitrary integration constant of dimension length.
Using
 these solutions to
calculate Einstein's tensor, we find the energy-momentum tensor
\begin{equation}
2\pi G \langle T^{t}_{\;\;t}\rangle =  \frac{\ell}{r^3}.
\end{equation}
Thus, the $1/r^3$-dependence of the energy-momentum tensor
follows from its algebraic form
when we use Einstein's equations. Because of the Bianchi identity,
energy momentum conservation also
holds in the quantum corrected geometry.
The integration constant $\ell$ is determined by the
vacuum energy density of equation (\ref{vacuum}) as
\begin{equation}
\ell =2\pi G a .
\end{equation}
Note that since the Casimir energy density,
$\rho=-a/r^3$, is negative, $\ell$ is
positive.
Solving equation (\ref{req}) for $u$ we get
\begin{equation}
1+2u = \left(1-\frac{2\ell}{r}\right)^{-1}
\end{equation}
which gives us the metric
\begin{equation}
ds^2=-\left(1-\frac{2\ell}{r}\right)dt^2+\left(1-\frac{2\ell}{r}
\right)^{-1}dr^2+
C_{0}^2 r^2d\theta^2 .
\end{equation}
This is the equatorial section of the (3+1)-dimensional
Schwarzschild metric
minus a wedge.
The Schwarzschild metric has a horizon which would pose a serious
problem of interpretation
if this were the exact solution of the
backreaction problem.
But here
one could argue that one should expand the metric
and keep only first order terms in $\ell$, since
a semi-classical approximation is not believed to
be valid
when the perturbations of the metric become of order unity.
To this it could be objected that
the expansion in Schwarzschild mass
is coordinate dependent: If one uses Eddington coordinates
the {\em exact\/} solution is already of first order in $\ell$.
Instead one could require the corrections to
the curvature scalars to be bounded by the Planck curvature.
Then one finds the constraint $r^3\gg\ell_{Pl}^2\ell$.
Now, the global geometry of the (2+1)-dimensional universe
is conical only if the sum of the angle deficits of all the particles is
less than $2\pi$ \cite{DeserSteif}. One might therefore argue that a single
particle should have a mass much smaller than the total mass of the universe,
and consequently that its angle deficit is much less than $2\pi$.
In this case \cite{SS}
$\ell\ll\ell_{Pl}$, and the curvature constraint implies that only
the region $r\gg\ell$ is well described by semi-classical Einstein theory.
In this way,
according to both cut off schemes the semi-classical approximation breaks down
before the horizon is reached. If particles are
to be described semi-classically, they must have radii
much larger than $\ell_{Pl}$, and then the
horizon problem is avoided.

Test particles in this geometry will feel an
inverse square law of gravitational attraction corresponding to
the gravitational potential\footnote{
Souradeep and Sahni \cite{SS} have obtained the same result
in the linearized approximation to the semi-classical Einstein equations.
They also calculate $\langle T^{\mu}_{\;\;\nu}\rangle$ explicitly.
The results
are in agreement with the symmetry considerations presented here. }
\begin{equation}
\Phi=-\frac{\ell}{r} .
\end{equation}
Despite its appearance, this is {\em not\/} the Newtonian potential.
Newton's theory gives a logarithmic potential
\cite{Giddinsetal} in 2+1 dimensions.
On the other hand, it is something of a paradox
that by adding a quantum vacuum having a negative energy density
to a particle which in the classical
limit produces no gravitational field, we get a {\em positive
gravitational mass\/}
and gravitational {\em attraction.\/}

\section{Resolving the positive mass problem}
By assuming a point particle at the origin one encounters
the paradox that
negative energy gives positive mass. Moreover, the vacuum polarization energy
diverges at the origin.
Since a point particle is a {\em singularity\/}
where space-time curvature diverges, it should not be
surprising that such a model gives unphysical results.
To avoid this problem one has to allow for a finite radius of the
source. Note that the exterior geometry remains the same, so
at distances much larger than $\ell$  one expects the same quantum field.

With a finite source it is clear that
one must take into account the
effect of vacuum polarization not only in the exterior of the particle
but also in its interior.
In connection with cosmic strings this point was stressed by
Frolov et al.\ \cite{Froloval}
who argued that ``it is not permissible to ignore the \ldots
interior of the string \ldots when assessing the
integrated effect of vacuum polarization''.
In the case of a cosmic string
we also have negative vacuum energy outside the source, yet
there is an attractive gravitational force there
[12-14].

In general it would be difficult to calculate this effect from first
principles,
but in this case we know the unique exterior
solution to Einstein's field equations.
Therefore, we may try to match the exterior solution to an interior one
corresponding to the classical source. The misfit may be identified with
a singular string on the junction according to Israel's
formalism \cite{Israel},
and in turn one may
identify this extra term in the energy-momentum
tensor with the integrated effect of vacuum polarization in the interior
\cite{GronSol}.

By matching the exterior solution
\begin{equation}
ds^2=-\left(1-\frac{2\ell C_{0}}{\rho}\right) dt^2+
\left(1-\frac{2\ell C_{0}}{\rho}\right)^{-1}\frac{d\rho^2}{C^2_{0}}
+\rho^2 d\theta^2
\end{equation}
where $\rho\equiv C_{0} r$ to the interior solution of Giddings
et al.\ \cite{Giddinsetal}
\begin{equation}
ds^2=-dt^2+(1-\lambda \rho^2 )^{-1} d\rho^2 + \rho^2 d\theta^2 ,
\end{equation}
where $\lambda$ is the energy-density,
and using Israel's formulae \cite{Israel},
the energy-momentum density of the junction string
is
given by the discontinuity of the exterior curvature at the junction.

Defining
\begin{equation}
K_{ij}\equiv -\frac{1}{2}\frac{d}{dR}g_{ij}
\;\;\;
{\mbox{ where }}
\;\;\;
dR^2\equiv g_{\rho\rho}d\rho^2 ,
\end{equation}
the Lanczos string energy-momentum tensor is
\begin{equation}
S^{i}_{\;j}=[K^{i}_{\; j}]-\delta^{i}_{\; j}[K]
\end{equation}
where
the square brackets signify the discontinuity at the junction string.
It follows that the string mass per length, $\mu$, and the string tension,
$\tau$, are
\begin{equation}
2\pi G\mu = [K^{\theta}_{\;\;\theta}]\;\;\;{\mbox{ and }}\;\;\;
2\pi G\tau =    [K^{t}_{\;\;t}] .\label{mutau}
\end{equation}
Using that
$C_{0}= (1-\lambda\rho_{0}^2)^{1/2}$ where $\rho_{0}$ is the radius of
the particle, and $\ell=2\pi G a$,
the mass per length and the tension of the junction string
derived from equations (\ref{mutau}) are
\begin{equation}
\mu=-\tau=\frac{a}{r^{2}_{0}}
\end{equation}
to first order in $\ell$.
Because the junction energy is identified with
the integrated effect of the vacuum polarization energy
of a massless
conformally coupled field,  it is
reasonable that the string energy-momentum is traceless.

In (3+1)-dimensional gravitation theory Tolman's \cite{Tolman}
gravitational
 mass is a very useful tool
to understand gravitational effects. In order
to derive a formula for
the (2+1)-dimensional effective gravitational mass, let us
in analogy with the four dimensional case \cite{Gron},
 consider
the acceleration of gravity, $k^{i}$,
as measured with standard rods and {\em coordinate clocks\/}
\begin{equation}
k^{i}=-\sqrt{-g_{tt}}\,\Gamma^{\hat{\imath}}_{\;\hat{t}\hat{t}} .
\end{equation}
Here, $\Gamma$ represents the connection coefficient and
hats indicate that it refers to an orthonormal frame.
The square root of the metric coefficient is introduced in order to
measure acceleration with coordinate clocks.
For a metric of the form (\ref{metrikk}) with $C=$constant,
 the acceleration is
radially directed and given by
$k=-A'/B$.
By use of the (2+1)-dimensional Einstein tensor
(\ref{ett})-(\ref{ethth}),  one finds for
continuously differentiable metrics that
\begin{equation}
k=-\frac{1}{r}\int_{0}^{r}
 \left(G^{r}_{\;\;r}+G^{\theta}_{\;\;\theta}\right)\sqrt{-g}dr .
\label{kdef}
\end{equation}
By relating the gravitational attraction to a concept of
gravitational mass,
 $M$,
one may {\em define\/}
$GM\equiv-kr$. Substituting $-GM/r$ for $k$ and
using Einstein's equations to substitute the
energy-momentum tensor
for $G^{\mu}_{\;\;\nu}$ in equation (\ref{kdef}), we derive the
 the following equation
\begin{equation}
M(r)= \int_{0}^{2\pi}\int_{0}^{r}
\left( T^{r}_{\;\;r}+T^{\theta}_{\;\;\theta}\right)
\sqrt{-g}drd\theta .
\end{equation}
This is
a (2+1)-dimensional analogue to Tolman's mass formula.
Since $T^{t}_{\;\;t}$ is missing from this expression,
 it is clear that (2+1)-dimensional gravity has no
Newtonian limit.

In the present case, there is a gravitational mass contribution
both from the
exterior quantum vacuum and from the junction shell which represents
the integrated effect of vacuum polarization in the interior of the particle.
To first order in the angle deficit and $\ell$,
the mass contribution from the exterior
Casimir energy is\footnote{Since
$\langle T^{\mu}_{\;\;\nu}\rangle$ is of first order in $\ell$,  we
may neglect the angle deficit
in $\sqrt{-g}$ in the first order approximation.}
\begin{equation}
M_{C}(r)=\int^{2\pi}_{0}\int^{r}_{r_{0}}
\left(\langle T^{r}_{\;\;r}\rangle +
\langle T^{\theta}_{\;\;\theta}\rangle \right)rdrd\theta
=2\pi \left[\frac{a}{r}-\frac{a}{r_{0}}\right].
\end{equation}
Note that $M_{C}(r)<0$ for all $r>r_{0}$, i.e.\ outside the particle.
The junction string has the effective mass
\begin{equation}
M_{J}=\int^{2\pi}_{0}(-\tau)r_{0}d\theta=
\frac{2\pi a}{r_{0}}
\end{equation}
which is a positive mass.
Physically we associate the positive mass of the junction string with
the integrated effect of vacuum polarization in the interior of the particle.
Adding these two terms we get the total vacuum polarization mass
\begin{equation}
M(r)= M_{C}(r)+M_{J}
= \frac{2\pi a}{r} .
\end{equation}
Hence, the gravitational potential, $\Phi=-\ell/r$, becomes
\begin{equation}
\Phi=- GM(r).
\end{equation}
This expression is Newtonian only by appearance, since
$M$ is a purely relativistic mass. Note also that $M$ is a function of $r$.

\section{Concluding remarks}
Although Einstein's theory in three-dimensional space-time
does not give Newtonian gravitation as a weak field limit,
it is amusing that quantum corrections and relativistic effects
give an attractive inverse square law
of gravitation. Note, however, that this
 also is a non-Newtonian law because in three dimensional space-time
Newtonian gravitation implies a logarithmic potential.

In connection with straight
cosmic strings it has been shown \cite{LT}
that linearized $R+R^2$ gravity produces short range
gravitational forces not present in pure Einstein theory.
It was argued that such forces might have had an effect on the formation of
structures in the early universe \cite{LT}.
It is therefore interesting that such effects appear also
in semi-classical Einstein theory, although the quantum
corrections to cosmic strings in Einstein's theory
may be too small to produce significant cosmological effects
\cite{GronSol}.

It is important to realize that the semi-classical corrections cannot
be self-consistently treated in the point mass approximation. In view of
the pathological properties of point masses,
it may be unfortunate that just point masses have been
given so much attention in
(2+1)-dimensional gravitation theory.
\subsection*{Acknowledgements}

This research was supported in part by the
Fridtjof Nansen Foundation Grant No.\ 152/92, by
Lise and Arnfinn Heje's Foundation Reference No.\ 0F0377/92,
the Norwegian Research Council for Science and the Humanities (NAVF),
Grant No.\ 420.92/022
and by
U.~S.\ DOE under Grant No.\ DOE-EY-76-C-02-3071.


\end{document}